\documentclass[superscriptaddress,aps,pre,twocolumn,floatfix]{revtex4-1}
\usepackage[utf8]{inputenc}
\usepackage{amsmath}
\usepackage{amssymb}
\usepackage{refcount}
\usepackage{siunitx}
\usepackage{graphicx}
\usepackage{hyperref}
 \usepackage[capitalise]{cleveref}
\usepackage{microtype}
\usepackage{bm}
\usepackage[english]{babel}
\usepackage{color}
\usepackage{graphicx}
\usepackage{xcolor}
\usepackage{braket}

\newcommand{\comm}[2]{\left[ #1 , #2 \right]}

\newcommand{\abs}[1]{\left| #1 \right|}

\newcommand{\br}{\mathbf{r}}

\newcommand{\bp}{\mathbf{p}}

\newcommand{\bE}{\mathbf{E}}

\begin{document}
\author{Haidar Al-Naseri}
\email{haidar.al-naseri@umu.se}
\affiliation{Department of Physics, Ume{\aa} University, SE--901 87 Ume{\aa}, Sweden}


\author{Gert Brodin}
\email{gert.brodin@umu.se}
\affiliation{Department of Physics, Ume{\aa} University, SE--901 87 Ume{\aa}, Sweden}
\title{Linear pair creation damping of high frequency plasma oscillation}
\pacs{52.25.Dg, 52.27.Ny, 52.25.Xz, 03.50.De, 03.65.Sq, 03.30.+p}

\begin{abstract}
We have studied the linear dispersion relation for Langmuir waves in plasmas of very high density, based on the Dirac-Heisenberg-Wigner formalism. The vacuum contribution to the physical observables leads to ultra-violet divergences, that are removed by a charge renormalization. The remaining vacuum contribution is small, and is in agreement with previously derived expressions for the time-dependent vacuum polarization. The main new feature of the theory is a damping mechanism similar to Landau damping, but where the plasmon energy give rise to creation of electron-positron pairs. The dependence of the damping rate (pair-creation rate) on wave-number, temperature, and density is analyzed. Finally, the analytical results of linearized theory are compared with numerical solutions.     
   
\end{abstract}
 
\maketitle

\section{Introduction}
As is well-known, increasing the number density in plasmas gradually introduces quantum phenomena in the picture. This has been explored both in older literature (see e.g. Ref. \cite{old-1}) as well as in many recent works (e.g. Refs. \cite{Manfredi-2019,Shukla-Eliasson-2011,Melrose-2020,Shukla-Eliasson-2011-2,Vladimirov-review,Own-review}). These works have been motivated e.g. by applications in various fields such as e.g. spintronics \cite{Spintronics}, plasmonics \cite{Plasmonics}, and astrophysics \cite{astro1,astro2}, where dense objects such as white dwarf stars or neutron stars are of particular interest. In the non-relativistic regime, replacing the Vlasov equation with the Wigner-Moyal equation \cite{Manfredi-2019,Shukla-Eliasson-2011,Melrose-2020,Shukla-Eliasson-2011-2,Vladimirov-review,Own-review}) covers the particle dispersive behavior, which is a key feature in many cases. More elaborate kinetic models also includes electron spin dynamics \cite{Own-review}), exchange effects \cite{Egen-2019-exchange} and collisional contributions, where the latter is of particular significance in the strong coupling regime \cite{Ichimaru82,Bonitz2015} 

For sufficiently high plasma densities, the Fermi velocity becomes relativistic, in which case a quantum relativistic treatments become necessary. Although a fair number of quantum relativistic plasma treatments have been made over the years (see e.g. Refs. \cite{Manfredi-2019,Shukla-Eliasson-2011,Melrose-2020,Shukla-Eliasson-2011-2,Vladimirov-review,Own-review})), still there are many simple but fundamental problems that have not been fully covered. In particular, as far as we know,  an analysis of the kinetic dispersion relation for Langmuir waves based on the Dirac equation has not been made. However, it should be stressed that somewhat simplified treatments, based on the Klein-Gordon equation for electrons have been published in the literature \cite{Haas-KG}.

In the present paper, we analyze the kinetic dispersion relation based on the Dirac equation, which is derived using the Dirac-Heisenberg-Wigner (DHW) formalism \cite{Birula}. Although the contribution from vacuum polarization is small, it must be given special attention, since it is related to the need for a re-normalization, due to ultra-violet divergences stemming from the vacuum contribution to the Dirac field. It is found that our expression for the vacuum polarization agrees with previous works derived using different methods, see Ref.  \cite{Mameyev-1981}. 

With the possible exception of very short wavelengths (of the order of the Compton length $\lambda_c=\hbar /mc$ where $m$ is the electron mass), it is confirmed that the classical but relativistic Vlasov equation gives an accurate approximation for the real part of the wave frequency. However, a new type of damping mechanism is introduced in the DHW-theory for very high plasma densities, namely pair-creation damping. Mathematically, the pair-creation resonances appear in the same way as the ordinary Landau pole, but it is only present providing the condition  $\hbar\omega_p>2mc^2$ is fulfilled, where $\omega_p$ is the plasma frequency. 

In addition to the analytical solution describing pair-creation damping, the DHW evolution equations are solved numerically. This confirms that wave-damping occurs due to pair-creation in the region of momentum space fulfilling the resonance condition. The scaling of the pair-creation damping with density (chemical potential), temperature, and wave-number are computed. Our work ends with a concluding discussion.    

\section{The DHW-formalism}

\subsection{The DHW equations}
In this sub-section, we will give a brief outline of the DHW-equations, to put the theory into context. For more detailed derivations, see the original source of the DHW-equations \cite{Birula} or the equivalent but slightly different presentations given in Refs. \cite{Gies2,Gies}.
We start with the gauge-invariant Wigner transformation
\begin{multline} \label{wigtrans}
\hat{W}(\br,\bp,t) = 
\\
\int d^3z
\exp \bigg(-i\textbf{p}\cdot\textbf{z} -ie\int^{1/2}_{-1/2} d\lambda \mathbf{z}\cdot \mathbf{A}(   \br+\lambda\mathbf{z},t )\bigg) 
\\ 
\times \hat{C}(\br,\bp,t),
\end{multline}
where
\begin{equation}
\label{Dirac-operatos}
    \hat{C}(\br,\bp,t) = - \frac{1}{2}
\comm{ \hat{\Psi}(\br+\mathbf{z} /2,t )} { \hat{\bar{\Psi}} (\br -\mathbf{z}/2,t )}, 
\end{equation}
$\hat{\Psi}$ is the four spinor of the Dirac theory, the bar denotes Hermitian transpose, and the bracket is the commutator.  
In \cref{wigtrans} we have used the Wilson line factor in the exponent (the contribution from the vector potential ${\bf A}$) to ensure gauge invariance.
The Wigner function $W(\br,\bp,t)$ is defined as the expectation value of the Wigner operator
\begin{equation}
\label{Wigner_function}
    W(\br,\bp,t)=\braket{\Omega |\hat{W}(\br,\bp,t)|\Omega}, 
\end{equation}
where $\ket{\Omega}\bra{\Omega}$ is the state of the system. 

The evolution of the Wigner function is computed by using the Dirac equation for $\hat{\Psi}$, where the electromagnetic field is treated classically, i.e. the fluctuations of the electromagnetic field are neglected. This approximation corresponds to ignoring higher-loop radiative corrections. It has been argued \cite{Birula} that the validity of this approximation may still allow field strengths up to or exceeding the Scwinger critical field, and also allow for short spatial scale lengths (of the order of the Compton length). However, as suggested in Ref. \cite{Birula} the theory may still have limitations regarding the temporal variations of the fields. We will address the issue of temporal limitations in \cref{Validity of DHW}. Here we just present our conclusion: The mean-field approximation used when deriving the DHW-equations may still allow for rapid variations of the electromagnetic field (of the order of the Compton frequency $\omega_c=mc^2/\hbar$), provided the plasma frequency is high enough, that is that the plasma frequency is comparable to the temporal scale of the EM-field. 

For the rest of this section, we put $\hbar=c=1$ to simplify the notation. After the Dirac equation and the mean field approximation have been applied, we obtain evolution equations for the 16 components of the Wigner function. To elucidate the physics of these equations, we expand the Wigner function $W(\br,\bp,t)$ in terms of an irreducible set of $4 \times 4$ matrices $\{\textbf{1},\gamma_5,\gamma^{\mu},\gamma^{\mu}\gamma_5,\sigma^{\mu,\nu} \}$ where $\textbf{1}$ is a $4\times 4$-identity matrix. We can then write the Wigner function as
\begin{equation}
\label{Expansion}
    W(\br,\bp,t)= \frac{1}{4}\Big[ s+ i\gamma_5 \varrho + \gamma^{\mu} v_{\mu}+ \gamma^{\mu}\gamma^5 a_{\mu} + \sigma^{\mu \nu }t_{\mu \nu}
    \Big],
\end{equation} 
 where the expansion coefficients $\{s,\varrho,v_{\mu},a_{\mu},t_{\mu\nu}\}$ are called the DHW-functions. By using this expansion of the Wigner function, dividing the four vectors $v_{\mu}$ and $a_{\mu}$ into their temporal ($v_0$ and $a_0$) and spatial (${\bf v}$ and ${\bf a}$) components, and splitting the anti-symmetric tensor $t_{\mu\nu}$ into the three-vectors ${\bf t}_1$ and ${\bf t}_2$, we finally obtain the DHW-equations \cite{Birula}:

\begin{align}
\label{DHW_System_diff}
    D_t s-2\Tilde{\textbf{p}}\cdot \textbf{t}_1&=0 \notag\\
    D_t\varrho+2 \Tilde{\textbf{p}}\cdot \textbf{t}_2 &=2ma_0\notag\\
    D_tv_0 +\textbf{D} \cdot \textbf{v}&=0\notag \\
    D_ta_0 + \textbf{D}\cdot \textbf{a}&=-2m \varrho  \\
    D_t \textbf{v} + \textbf{D}v_0 - 2\Tilde{\textbf{p}}\times \textbf{a}&=-2m\textbf{t}_1\notag \\
    D_t\textbf{a} + \textbf{D}a_0 - 2\Tilde{\textbf{p}}\times \textbf{v}&=0\notag \\
    D_t\textbf{t}_1+ \textbf{D}\times \textbf{t}_2 + 2\Tilde{\textbf{p}} s&=2m\textbf{v} \notag \\
    D_t \textbf{t}_2- \textbf{D}\times \textbf{t}_1 
    -2\Tilde{\textbf{p}}\varrho &=0.\notag 
\end{align}
where we have introduced the non-local operators
\begin{align}
D_t&= \frac{\partial}{\partial t} + e\Tilde{\bE}\cdot \bm\nabla_p\\
    \Tilde{\textbf{p}}&=\textbf{p}- ie\int^{1/2}_{-1/2}d\tau \tau \textbf{B}(\br+i \tau \bm \nabla_p)\times \bm\nabla_p\\
    \textbf{D}&= \nabla_r+ e\int^{1/2}_{-1/2}d\tau \tau \textbf{B}(\br+i\tau \bm \nabla_p)\times \bm \nabla_p\\
    \Tilde{\textbf{E}}&= \int^{1/2}_{-1/2}d\tau  \textbf{E}(\br+i \tau \bm\nabla_p) 
\end{align}
 Most of the DHW-functions have a clear physical interpretation, as will be briefly outlined below. Firstly, the electromagnetic current $J^{\mu}$ can be expressed as
\begin{equation}
J^{\mu}= \frac{e}{(2\pi )^3} \int d^3p\, v^{\mu} (\br,\bp,t)
\end{equation}
where the total charge Q is
\begin{equation}
\label{Conservation_charge}
    Q=\frac{e}{(2\pi)^3}\int d^3pd^3r  v_0(\br,\bp,t)
\end{equation}
Moreover, the total energy $W$ is given by
\begin{multline}
\label{Conservation_Energy}
W= \frac{1}{(2\pi )^3}\int d^3pd^3r \big[\mathbf{p}\cdot \mathbf{v}(\br,\bp,t) + ms(\br,\bp.t)   \big]\\
+ \frac{1}{2} \int d^3r \big[E^2+B^2   \big].
\end{multline}
The linear momentum is 
\begin{equation}
\label{Momentum}
    \textbf{p}= \frac{1}{(2\pi )^3}\int d^3pd^3r\, \textbf{p} v_0(\br,\bp,t)  + \int d^3r \textbf{E}\times \textbf{B}
\end{equation}
and the total angular momentum $\textbf{M}$ is 
\begin{multline}
\label{Angular_Mom}
    \textbf{M}= \frac{1}{(2\pi )^3}\int d^3pd^3r \Big[\textbf{r}\times \textbf{p} v_0(\br,\bp,t) + \frac{1}{2}\textbf{a}(\br,\bp,t)  \Big]\\ + \int d^3r \, \textbf{r}\times \textbf{E}\times \textbf{B}
\end{multline}
The interpretation that can be done from the expressions above is that $s(\br,\bp,t)$ is the mass density, $v_0(\br,\bp,t)$ the charge density and $\mathbf{v}(\br,\bp,t)$ is the current density. Moreover, as seen from Eq. (\ref{Angular_Mom}), the function $\mathbf{a(\br,\bp,t)}$ can be associated with the spin density. For further elaborations regarding the interpretations of the DHW-functions, see e.g. Refs \cite{Birula, Review}

The classical, but still relativistic, Vlasov equation can be obtained in the limit $\hbar \rightarrow 0$ (which would require us to first reinstate $\hbar$ in the governing equations, see e.g. Ref. \cite{Birula}). 
Note, however, that the variable $v_0$, which is proportional to the charge density, must be kept non-zero. 
Thus the procedure to reach the classical limit, which is outlined in Ref.~\cite{Birula}, must be somewhat modified, see Ref. \cite{Review} for a detailed discussion.

\subsection{One-dimensional electrostatic fields}


In this subsection, we simplify the DHW-system  by considering one-dimensional electrostatic fields, $\bE(t,\br)=E(t,z)\textbf{e}_z$. In general, we can express the DHW-function in terms of new dependent variables $\chi_i(z,\bp,t)$ as
\begin{equation}
    \{s,  \varrho ,  v_{\mu}, a_{\mu} ,\mathbf{t}_1,\mathbf{t}_2 \}=\sum_{i=1}^{16} \chi_i(z,\bp,t)\textbf{e}_i(z,\bp,t).
\end{equation}
With a suitable choice of eigenvectors $\textbf{e}_i(z,\bp,t)$, it can be shown that in the limit of consideration, only four of the 16 variables of $\chi_i$ are linearly independent. For the specific choice of eigenvectors $\textbf{e}_i(z,\bp,t)$ made in Ref.  \cite{PRE21}, the DHW-system \cref{DHW_System_diff}  reduces to
\begin{align} 
\label{PDE_System}
    D_t\chi_1(z,\bp,t)&= 2\epsilon_{\bot}(p_{\bot}) \chi_3(z,\bp,t)- \frac{\partial \chi_4}{\partial z} (z,\bp,t)\notag\\
    D_t\chi_2(z,\bp,t) &= -2p_z\chi_3(z,\bp,t)\\
    D_t\chi_3(z,\bp,t)&= -2\epsilon_{\bot}(p_{\bot}) \chi_1(z,\bp,t) +2p_z\chi_2(z,\bp,t)\notag\\
    D_t\chi_4(z,\bp,t)&= -\frac{\partial \chi_1}{\partial z}(z,\bp,t)\notag 
\end{align}
where we have introduced $\epsilon_{\bot}=\sqrt{m^2+p_{\bot}^2}$.
This system of four coupled equations is closed by Ampére's law
\begin{equation}
\label{Ampers_law}
\frac{\partial E}{\partial t}=-\frac{e}{(2\pi)^3}\int \chi_1 d^3p
\end{equation}
where we have used the relation between the original DHW-functions and the the expansion functions $\chi_i(z,\bp,t)$. The complete list of relations between the two sets of variables are as follows:
\begin{align}
\label{DHW_Non}
    s(z,\bp,t)&= \frac{m}{\epsilon_{\bot}}\chi_2(z,\bp,t) \notag\\
    v_0(z,\bp,t)&= \chi_4(z,\bp,t)\notag \\
    \textbf {v}_{\bot}(z,\bp,t)&=\frac{ \textbf{p}_{\bot}}{\epsilon_{\bot}}\chi_2(z,\bp,t) \notag \\
    v_z(z,\bp,t)&=\chi_1(z,\bp,t) \\
    a_x(z,\bp,t)&=-\frac{p_y}{\epsilon_{\bot}}\chi_3(z,\bp,t)\notag\
    a_y(z,\bp,t)&=\frac{p_x}{\epsilon_{\bot}}\chi_3(z,\bp,t)\notag\\
    t_{1z}(z,\bp,t)&=- \frac{m}{\epsilon_{\bot}} \chi_3(z,\bp,t) \notag 
\end{align}
Evidently, $\chi_1$ is closely related to the phase-space current density parallel to the field, $\chi_2$ to the mass density (or the perpendicular current density), $\chi_3$ to the spin density, and $\chi_4$ to the charge density. 

\subsection{Linear Waves}
In what follows, we will study plane wave Langmuir waves in a homogeneous electron-positron ion plasma, using linearized theory. The ions will not be treated dynamically, but will constitute a constant neutralizing background. In the rest of this work, we will reinstate $\hbar$ in order to distinguish classical and quantum terms. 
Now, in order to study linear theory, we divide the functions $\chi_i$ into background functions $\chi_i^0$ and perturbed functions $\chi_i^1$
\begin{equation}
\chi _{i}(z,\bp,t)=\chi _{i}^{0}(\bp)+\chi _{i}^{1}(\bp)e^{i(kz-\omega t)}
\end{equation}%
To find proper expressions for the background functions $\chi_i^0(\bp)$, we first notice that
the nonzero vacuum expectation values contribute to the background quantities $\chi _{i}^0 (\bp)
$ through the current density $s(\bp)$ and mass density $\mathbf{v}(\bp)$ as 
\begin{align}
s_\text{vac}(\bp)& =-\frac{2m}{\epsilon }  \notag  \label{Vacuum_sol} \\
\mathbf{v}_\text{vac}(\bp)& =-\frac{2\mathbf{p}}{\epsilon },
\end{align} 
see e.g. \cite{Birula}.
Furthermore, using the relations between the DHW-variables and $\chi_i$-variables in \cref{DHW_Non}, we get
\begin{align}
\label{Vacuum_contribution}
\chi _{1}^0(\bp)& =-\frac{2p_{z}}{\epsilon }  \notag \\
\chi _{2}^0(\bp)& =-\frac{2\epsilon _{\bot }}{\epsilon }.
\end{align}%

Moreover, a background distribution
function $f_{e}(\bp)$ of electrons ($f_{p}(\bp)$ for positrons), can be added to the vacuum background as
follows: 
\begin{align}
v_{0}& =2(F+1) \\
s(\bp)& =\frac{2m}{\epsilon }F(\bp) \\
\mathbf{v}(\bp)& =\frac{2\mathbf{p}}{\epsilon }F(\bp),
\end{align}%
where $F(\bp)=[f_{p}(\bp)+f_{e}(\bp)-1]$. In terms of the new functions $\chi _{i}$, we have 
\begin{align}
\label{Initial_values}
\chi _{1}^{0}(\bp)& =\frac{2p_{z}}{\epsilon }\Big[f_{p}(\bp)+f_{e}(\bp)-1\Big] 
\notag \\
\chi _{2}^{0}(\bp)& =\frac{2\epsilon _{\bot }}{\epsilon }\Big[%
f_{p}(\bp)+f_{e}(\bp)-1\Big] \\
\chi _{4}^{0}(\bp)& =2\Big[f_{p}(\bp)-f_{e}(\bp)\Big]  \notag
\end{align}%
The electron/positron background distribution function $f_e(\bp)/f_p(\bp)$ are normalized
such that the unperturbed number density $n_{0}$ is
\begin{equation}
\label{normalization}
    n_{0e,p}=\frac{2}{(2\pi 
 \hbar)^{3}}\int f_{e,p}(\bp)d^{3}p,
\end{equation}
Without ions contributing to the charge density, as written above, we must have a neutral electron-positron background (i.e. $n_0=n_{0e}=n_{0p}$). Adding an ion species and letting the electron and positron densities background densities differ is trivial, however.  

The function  $f_{e,p}(\bp)$ can be picked as any common background distribution function from classical kinetic theory, i.e. a Maxwell-Boltzmann, Synge-Juttner, or Fermi-Dirac distribution, depending on whether the characteristic kinetic energy is relativistic and whether the particles are degenerate. In what follows, we will put the initial positron density to zero, and consider a partially or completely degenerate Fermi-Dirac electron background. 
Note that for a
completely degenerate ($T=0$) Fermi-Dirac background of electrons (and no positrons $f_{p}=0$), the electron and vacuum contributions cancel  inside the Fermi sphere. Consequently, for
momenta $p\leq p_{F}$, where  $p_{F}=\hbar (3\pi^2n_{0})^{1/3}$ is the Fermi
momentum we have $F(\bp)=0$. 
 
 Before staring the linear analysis, we first note a helpful relation regarding the non-local operator contained in ${\bf \tilde{E}}$, namely  
\begin{equation}
\Tilde{\mathbf{E}}\cdot \nabla _{p}\chi _{i}^{0}=\tilde{E}\frac{\partial
\chi _{i}^{0}}{\partial p_{z}}=E\frac{\chi _{i}^{0}(p_{z}+\hbar k/2)-\chi
_{i}^{0}(p_{z}-\hbar k/2)}{\hbar k}.  \label{Derivative generalization}
\end{equation}%
With this relation established, the linear theory is now reduced to standard linear algebra.  Solving for $\chi _{i}^{1}(\bp)$ in terms of $\chi_i^0(\bp)$ we
obtain
\begin{widetext}

\begin{small}
    
\begin{align}
    \chi_1^1(\bp)&=   \sum_{\pm}
    \frac{\pm i2e\omega E/ (\hbar k) }{(\omega^2-k^2)(\hbar^2\omega^2-4p_z^2)-4\epsilon_{\bot}^2\omega^2}
    \Bigg[ 4p_z\epsilon_{\bot}^2   \frac{F(p_{\pm})}{\epsilon_{\pm}}  
    -(\hbar^2\omega^2 -4p_z^2)
    \bigg( \frac{p_{\pm}}{\epsilon_{\pm}} F(p_{\pm})
  + \frac{k}{\omega}\Big(f_p(p_{\pm})-f_e(p_{\pm})\Big)
    \bigg)
    \Bigg]
    \label{chi1} 
    \\
    \chi_2^1(\bp)&=  \sum_{\pm}
    \frac{\mp  i\omega eE \epsilon_{\bot } / (\hbar k) }{(\omega^2-k^2)(\hbar^2\omega^2-4p_z^2)-4\epsilon_{\bot}^2\omega^2}
    \Bigg[  \Big( \hbar^2 \omega^2 -\hbar ^2k^2 - 4\epsilon^2 \mp \frac{\hbar k }{2}p_z\Big)   \frac{F(p_{\pm})}{\epsilon_{\pm}}  
  - 4p_z\frac{k}{\omega}\Big(f_p(p_{\pm})-f_e(p_{\pm})\Big)
    \Bigg]\\
    \chi_3^1(\bp)&= \sum_{\pm}
    \frac{\mp  4\omega eE \epsilon_{\bot }  }{(\omega^2-k^2)(\hbar^2\omega^2-4p_z^2)-4\epsilon_{\bot}^2\omega^2}
    \Bigg[  \Big( p_z\frac{k}{\omega}\pm \frac{\hbar \omega }{2}\Big)\frac{F(p_{\pm})}{\epsilon_{\pm}}   + f_p(p_{\pm})-f_e(p_{\pm})
    \Bigg]\\
    \chi_4^1(\bp)&= \sum_{\pm}
    \frac{\pm  2i\omega eE/(\hbar k)  }{(\omega^2-k^2)(\hbar^2\omega^2-4p_z^2)-4\epsilon_{\bot}^2\omega^2}
    \Bigg[  \big(4\epsilon^2-\hbar^2\omega^2  \big)
    \left[
    \frac{kp_z}{\omega}\frac{F(p_{\pm})}{\epsilon_{\pm}}  + f_p(p_{\pm})-f_e(p_{\pm})\right] 
    \pm \frac{\hbar k^2}{2\omega}
    \big(4p_z^2-\hbar ^2\omega^2   \big) \frac{F(p_{\pm})}{\epsilon_{\pm}} 
    \Bigg]
    \label{chi4}
\end{align}
\end{small}
\end{widetext} 
where
\begin{align}
p_{\pm}&= p_z\pm \frac{\hbar k}{2}\\
\epsilon_{\pm}&= \sqrt{m^2+p_{\bot}^2 + \Big(p_z \pm \frac{\hbar k }{2}\Big)^2}
\end{align}
Note that $F(p_{\pm})$ and $f_{e,p}(p_{\pm})$ depend on the full momentum, but we suppressed the perpendicular momentum to simplify the notation.
Combining the above results for $\chi_i(\bp)$ with Ampere's law \cref{Ampers_law} we obtain the dispersion relation $D(k,\omega)=0$ with
\begin{small}
    
\begin{multline}
\label{Dispersion relation}
    D(k,\omega)= 1+ \sum_{\pm} \int \frac{d^3p}{(2\pi\hbar)^3}
    \frac{\pm 2e^2 / (\hbar k) }{(\omega^2-k^2)(\hbar^2\omega^2-4p_{\pm}^2)-4\epsilon_{\bot}^2\omega^2}\times\\
    \Bigg[ 4\frac{\epsilon_{\bot}^2}{\epsilon}   p_{\pm} F(\bp)  
    -(\hbar^2\omega^2 -4p_{\pm}^2)
    \bigg( \frac{p_z}{\epsilon } F(\bp)
  + \frac{k}{\omega}\Big(f_p(\bp)-f_e(\bp)\Big)
    \bigg)
    \Bigg] 
\end{multline}
\end{small}
The expression (\ref{Dispersion relation}) was presented in a previous paper \cite{PRE21}, but not analyzed.  To our knowledge, a full kinetic dispersion relation for Langmuir waves based on the Dirac equation has not been studied previously. The closest results to compare with are quantum relativistic treatments where electrons have been described by the Klein-Gordon (KG) equation, see e.g. \cite{Haas-KG} and a similar semi-relativistic quantum model studied by Ref. \cite{Extra-rel}. In particular, Ref. \cite{Haas-KG} gives a thorough treatment, where the combined influence of relativistic and short-scale physics (quantum recoil) is analyzed. Since the Klein-Gordon (KG) equation leaves out spin effects, and the variable $\chi_3$ (that is proportional to the spin density) is nonzero in our case, there cannot be a precise agreement between Eq. (\ref{Dispersion relation}) and previous results. Still, the KG treatment correctly captures classical relativistic effects as well as particle dispersive behavior. Moreover, even the somewhat more exotic effects of Zitterbewegung (rapid oscillations due to interference between positive and negative energy states), and pair-creation are covered by the KG equation. In fact, a phase space description much like the DHW-formalism has been developed based on the KG equation, where the 16 scalar components of DHW-theory are replaced by only four components \cite{KGWigner}. 

However, while the KG treatment of Ref. \cite{Haas-KG} is likely to give a useful approximation of the real part of the Langmuir frequency in the quantum relativistic regime, in that work the vacuum contribution to the background quantities was omitted.  As we will study in a separate sub-section, though, the vacuum polarization contribution to the real part of the frequency is generally small, Nevertheless, the vacuum contribution is crucial for the highest plasma densities.  The reason is that the dispersion relation (\ref{Dispersion relation}) exhibits pair-creation resonances, leading to wave damping. Without including the vacuum contribution to the background variables, this process cannot be studied, since without the vacuum contribution, we will get the wrong sign of the imaginary part of the frequency, i.e. we will get wave growth instead of damping. This conclusion can be reached without going through all the technical details, since ${\rm Im}\omega$ will be determined by the sign of the integrand, and the sign of $F$ changes due to the vacuum contribution. 

Before we get on to analyze the quantum relativistic contributions, let us first establish the connection with the classical (but still relativistic) limit. Letting $\hbar \rightarrow 0$, the dispersion function (\ref{Dispersion relation}) reduces to
\begin{multline}
\label{class_limit}
    D(k,\omega )= 1+ \frac{e^2}{\omega } \int \frac{d^3p}{(2\pi\hbar)^3} \frac{p_z}{\epsilon}
    \bigg(\frac{1}{\omega-kp_z/\epsilon}+ \frac{1}{\omega+kp_z/\epsilon}
    \bigg)\times\\
    \bigg[ \Big(1 + \frac{kp_z}{\epsilon \omega}
    \Big)\frac{\partial f_p(\bp)}{\partial p_z}
    + \Big(1 - \frac{kp_z}{\epsilon \omega}
    \Big)\frac{\partial f_e(\bp)}{\partial p_z}
    \bigg]. 
\end{multline}

We note that the appearance of $\hbar$ in the integration measure $\frac{d^3p}{(2\pi\hbar)^3}$ is just a matter of normalization (compare \cref{normalization}), and not a sign of any remaining quantum features. For background distributions $f_p$ and/or $f_e$ that are  even functions of $p_z$, it is straightforward to show that expression (\ref{class_limit}) agrees with results based on the classical (but relativistic) Vlasov equation. However, in case there are nonzero odd contributions to the background distribution, at first there seem to be a notable difference. In particular, for the classical case, the dispersion relation should be the same if a background distribution of electrons is replaced by the same background distribution of positrons (since the end result only contains the square of the particle charge). However, in Eq. (\ref{class_limit}), we see that there is a momentum anti-symmetry between the contributions from electrons and positrons, since we must use ${\bf p}\rightarrow -{\bf p}$ in the integrand to get the same expression when replacing the electrons with positrons. This is a general feature of the DHW-formalism, where positrons can be viewed as electrons moving backward in time. In particular, for a non-relativistic non-degenerate Maxwellian particle beams of electrons, with a distribution $\propto \exp(-({\bf p}-p_b{\bf z})^2/p_t^2)$, we must use a distribution $\propto \exp(-({\bf p}+p_b{\bf z})^2/p_t^2)$ for a beam of positrons {\it moving in the same direction}. Once this property has been noted, which obviously is different from the classical case, it is straightforward to confirm the exact and general agreement of Eq. (\ref{class_limit}) with the result computed from the relativistic Vlasov equation.  

\section{Renormalization and vacuum polarization}
For large momentum, where the approximations $\epsilon\approx p$,  $\epsilon_{\perp}\approx p_{\perp}$, etc. apply, it is straightforward to confirm that the integrand in \cref{Dispersion relation} scales with momentum as $1/p$, that is, there is a logarithmic divergence in the integral occurring for large momentum  due to the vacuum contributions of $F$ (note that the vacuum part does not decline with momentum, in contrast to the real particle contribution). This is a well-known feature of the DHW-formalism \cite{Birula} and of quantum field theory in general, referred to as the ultra-violet divergences.  This lead to the standard ultraviolet charge renormalization, which must be performed before the full dispersion can be evaluated. 
Thus we need to renormalize the charge in order to absorb the divergent term. Firstly, we rewrite the dispersion relation in \cref{Dispersion relation} as
\begin{equation}
    D(k,\omega)=1 + \frac{e_B^2}{\epsilon_0}\eta_v + \frac{e_B^2}{\epsilon_0}\eta_p, \label{renorm}
\end{equation}
where $e_B$ is the bare charge and $\eta_{v/p}$ is the vacuum/plasma contribution to the dispersion relation in \cref{Dispersion relation}. Thus, by definition, in the factor $F$ of Eq. (\ref{Dispersion relation}), the constant vacuum term contributes to $\eta_{v}$ whereas the electron/positron backgrounds contributes to $\eta_{p}$ 

The vacuum contribution $\eta_v$, after integration, will give us a logarithmic divergent term and a finite term that will be the part of the vacuum polarization that remains after the renormalization.
To illustrate the main principle, we first consider the somewhat simplified case of modest frequencies and wave-numbers, i.e. we study the case of $\hbar\omega \ll \epsilon$ and $\hbar k \ll \epsilon$. Keeping the first order corrections in a Taylor-expansion for $\eta_v$ (cf. Eqs. \cref{Dispersion relation} and (\ref{renorm})), we obtain 

\begin{multline}
\label{Renorm}
\eta_v=
    \frac{1}{2(2\pi )^3\hbar}\int^{\lambda}_{0} d^3p \frac{1}{\epsilon^3}
    \Bigg[ 1-\frac{p_z^2}{\epsilon^2}
    -\frac{\hbar^2 k^2 }{8\epsilon^2}
    \bigg( 3-5\frac{p_z^2}{\epsilon^2}+7\frac{p_z^4}{\epsilon^4}
    \Bigg)\\
    +\frac{\hbar^2\omega^2}{4\epsilon^2}\bigg(1-\frac{p_z^2}{\epsilon^2}
    \bigg)
    \Bigg]
\end{multline}
where we have introduced a momentum cut-off $\lambda$ to get a finite expression. Solving the integrals, we get
\begin{equation}
\label{Mamev}
    \eta_v= \frac{1}{24\pi^2\hbar } \ln{\Big(\frac{\lambda}{m}\Big)} 
    -\frac{1}{60\pi^2 \hbar m^2}\Big(\hbar^2k^2-\hbar^2 \omega^2\Big) 
\end{equation}
The first term of $\eta_v$ is the logarithmic term that will be absorbed by picking a proper renormalized charge $e_r$, and the second term is the vacuum polarization. Letting the renormalized charge be given by
\begin{equation}
   e_r^2 = \frac{e_B^2} {1+\frac{e_B^2}{24\pi^2\hbar } \ln{\Big(\frac{\lambda}{m}\Big)}} 
\end{equation}
the logarithmic term is indeed absorbed, and the relation between $e_r$ and $e_B$ agrees with Ref. \cite{Bloch}.
Next, using the expression for the renormalized charge in \cref{renorm}, we get
\begin{equation}
    D(k,\omega)=1+\frac{e_r^2}{\epsilon_0} \bigg[
    \eta_p
        -\frac{\hbar^2k^2-\hbar^2 \omega^2}{60\pi^2 \hbar m^2}
    \bigg]
\end{equation}
The term proportional to $\eta_p$ contains the same contribution from a real background distribution of particles as before, except that the bare charge is replaced by the re-normalized charge. Moreover, we note that the vacuum polarization term (second term in the square bracket) agrees with a general expression (proportional to derivatives of the EM-fields) for the space- and time-dependent vacuum contribuion (i.e. proportional to derivatives of the EM-fields) derived in Ref. \cite{Mameyev-1981}.

However, for very high frequency oscillations,  we cannot treat $\hbar \omega /\epsilon$ as a small parameter, except for the tail end of momentum space. To study the high-frequency regime we will now generalize the renormalization procedure, but consider only $k=0$ for simplicity. For this case, the vacuum contribution in \cref{Dispersion relation} $\eta_v$ is given by

\begin{equation}
    \eta_v=- \frac{4}{(2\pi)^2\hbar}\int dp \frac{p^2 }{ \epsilon(\hbar^2\omega^2-4\epsilon^2)}
    \Big(1-\frac{p^2}{3\epsilon^2 } \bigg) 
\end{equation}
To single out the ultra-violet divergent term, we expand the denominator in powers of  $\hbar \omega /\epsilon$ for the part of momentum space where $\hbar \omega < \epsilon$, and in inverse powers for the opposite part of momentum space. We then obtain
\begin{multline}
 \eta_v= \frac{4}{(2\pi)^2\hbar}
 \int^{\infty}_{p_{res}}dp \frac{p^2}{\epsilon^3}    \bigg(1-\frac{p^2}{3\epsilon^2 } \bigg) 
    \sum_{n=0}^{\infty} \bigg(\frac{\hbar^2\omega^2}{4\epsilon^2}\bigg)^n\\
      -\frac{16}{\omega^2(2\pi)^2\hbar^3}\int_{0}^{p_{res}}dp \frac{p^2}{\epsilon}    \bigg(1-\frac{p^2}{3\epsilon^2 } \bigg) 
    \sum_{n=0}^{\infty} \bigg(\frac{\epsilon^2}{\epsilon_{res}^2}\bigg)^n
\end{multline}
where 
\begin{equation}
    \epsilon_{res}^2= m^2+p_{res}^2
\end{equation}
and where $p_{res}$ is the resonant momentum
\begin{equation}
\label{Resonant_mom}
    p_{res}= m\sqrt{\frac{\hbar^2 \omega^2 }{4m^2}-1}
\end{equation}
 The first term of the first summation, i.e. for $n=0$, is the logarithmically divergent term, while all other terms constitute the vacuum polarization contribution. Thus, for $k=0$, after the renormalization has been made, the dispersion relation can be written
 \begin{equation}
    D(k=0,\omega)=1+\frac{e_r^2}{\epsilon_0} \bigg[
    \eta_p+\eta_v
    \bigg]
\end{equation}
where, in this case, $\eta_p$ stands for the contribution from the real particles in the limit $k=0$, and the the vacuum polarization is given by
\begin{multline}
 \eta_v= \frac{4}{(2\pi)^2\hbar}
 \int^{\infty}_{p_{res}}dp \frac{p^2}{\epsilon^3}    \bigg(1-\frac{p^2}{3\epsilon^2 } \bigg) 
    \sum_{n=1}^{\infty} \bigg(\frac{\hbar^2\omega^2}{4\epsilon^2}\bigg)^n\\
      -\frac{16}{\omega^2(2\pi)^2\hbar^3}\int_{0}^{p_{res}}dp \frac{p^2}{\epsilon}    \bigg(1-\frac{p^2}{3\epsilon^2 } \bigg) 
    \sum_{n=0}^{\infty} \bigg(\frac{\epsilon^2}{\epsilon_{res}^2}\bigg)^n
    \label{etafunc}
\end{multline}
\begin{figure}
    \centering
    \includegraphics[width=\columnwidth]{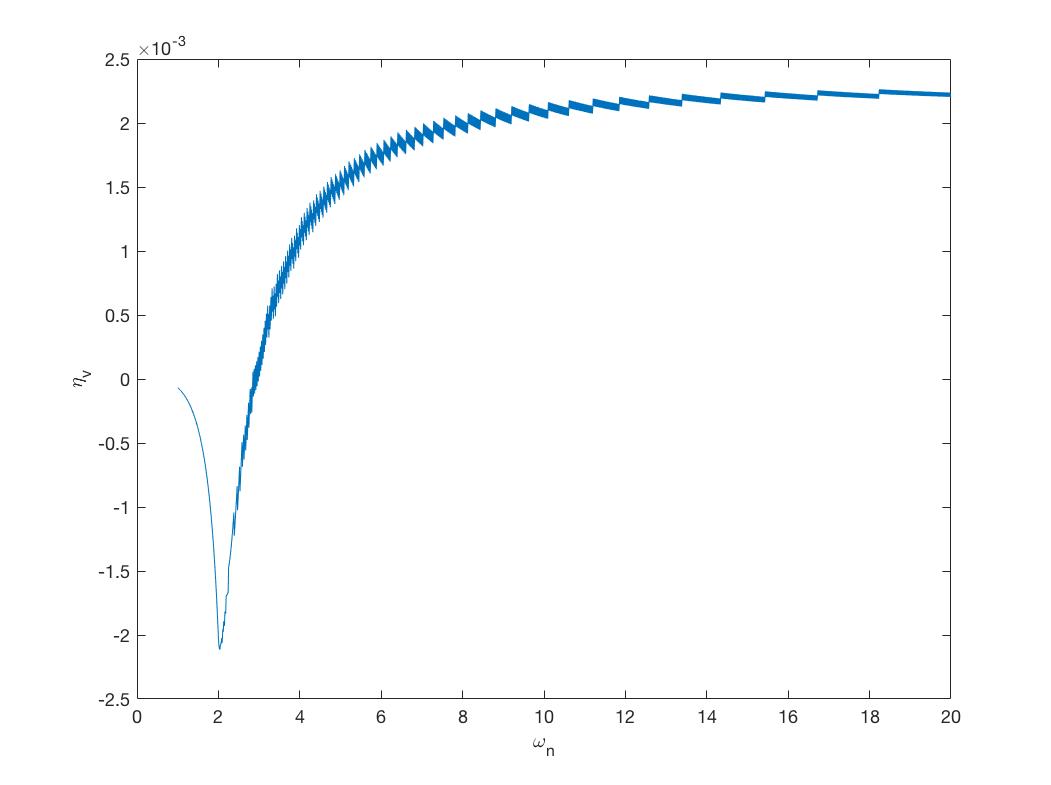}
    \caption{Vacuum polarization $\eta_v$ as a function of the normalized frequency $\omega_n$.}
    \label{eta_vs_omega}
\end{figure}
The full frequency dependence of the vacuum polarization given in \cref{etafunc} is plotted in \cref{{eta_vs_omega}}. A few things should be noted. Firstly, the overall magnitude of this contribution is always small, i.e. smaller than the contribution from the real particles by almost three orders of magnitude or more. Secondly, the quadratic increase with frequency is saturated for $\omega\approx 2\omega_c$, where the peak in the vacuum polarization is reached, a result which generalizes the findings of Ref \cite{Mameyev-1981}. Thirdly, we note that the curve exhibits short-scale oscillations. However, these oscillations are not a physical feature, but numerical noise associated with the truncation of the sum in Eq. (\ref{etafunc}) to a finite series.

While the vacuum-polarization is interesting from a theoretical point of view, the above findings suggest that due to the limited magnitude of this term, the vacuum polarization has little practical importance when wave propagation of Langmuir waves is concerned. However, so far we have deliberately avoided the issue of the pole contribution to the momentum integral, which should be included when the denominator becomes zero (i.e. when $\hbar\omega=2\epsilon$). As we will see in the next section, for high enough wave frequencies, the vacuum term is indeed crucial to obtain the correct pole contribution and thereby to compute the proper value of the wave damping.  
\section{Linear pair-creation-damping}

From now on
we will also include the effect of the  background distribution of electrons in \cref{Dispersion relation} together with the vacuum contribution. 
An analysis of the dispersion relation (\ref{Dispersion relation}) is helped by the observation that the quantum contribution to the {\it real frequency} is small, as long as the wavenumber is modest (well beyond the Compton wavelength). This is true even for large wave frequencies, of the order of the Compton frequency or higher. Thus, unless $k$ is very large, the relativistic Vlasov equation is a good approximation for the {\it real} part of the frequency $\omega_r$. The reason is that for $\hbar\omega\sim mc^2$, the Fermi energy $E_F$ will be much larger than unity. Thus, even if $\hbar\omega\sim mc^2$ we will have $\hbar\omega \ll \epsilon$, which, in turn, implies a minor quantum contribution to \cref{Dispersion relation}, not counting the short scale effects introduced by a large $k$. To be more precise, a plasma frequency of the order the Compton frequency (corresponding to $n_0\sim 3\times 10^{34} {\rm m^{-3}})$ gives $E_F/mc^2 > 100$. That is, for a modest $k$ the quantum terms in the dispersion relation (\ref{Dispersion relation}) only modify the real part of the frequency by a small correction of the order one percent  or so, as compared to the relativistic Vlasov equation. As we will see below, however, the same conclusion does not apply for the imaginary part of the frequency, $\omega_i$, where the full quantum relativistic theory is needed.     

To illuminate the physics of the pair creation damping, let us present some illustrative relations. Firstly, Taylor expanding the dispersion relation $D(\omega,k)=0$ for small relative damping (with $\omega=\omega_r + i\omega_i$, using $\omega_i\ll \omega_r$) we find that relative damping $\gamma$ simply is given by
\begin{equation}
    \gamma\equiv \frac{\omega_i}{\omega_r}=-\frac{{\rm Im}D(\omega_r,k)}{\omega_r(\partial D/\partial\omega_r)}
\end{equation}
where ${\rm Im}D(\omega_r,k)$ can be evaluated using the standard Landau contour when performing the momentum integration. From the relative damping, we can compute the pair creation rate, by using
\begin{equation}
dW_v/dt + dW_p/dt=0
\end{equation}
where 
\begin{align}
    W_v&= \epsilon_0\omega\frac{\partial D} {\partial \omega}\frac{\abs{E}^2}{2}\\
    W_p&=\int d^3p\, n \epsilon 
    \label{Particle_energy}
\end{align}
where $W_v$ is the energy density of the wave, $W_p$ is the energy density of the created pairs and $n$ is the number of the created pairs. Combining the above relations, we find 
\begin{equation}
    \frac{dW_p}{dt}=\int d^3p\, \frac{dn}{dt} \epsilon =\epsilon_0{\rm Im}D\,\omega_r\abs{E}^2
\end{equation}
Formally, the expression $\omega(\partial D/\partial \omega)$ is a rather complicated expression, but for the most part, it is rather close to 2, indicating that the energy density of the plasma wave is equally divided between electrostatic and kinetic degrees of freedom. Thus, both the relative wave damping $\gamma$ and the pair production rate is effectively determined by ${\rm Im} D$. Hence, our focus for the reminder of this section will be to determine the dependence of ${\rm Im} D$ on the basic parameters of the problem, i.e. the temperature, the chemical potential (electron number density) and the wavenumber. 
\subsection{Homogeneous limit}
 We will firstly consider the homogeneous limit, i.e. we let $k\rightarrow 0$ in \cref{Dispersion relation}. After taking the homogeneous limit we use spherical coordinates in momentum space and first perform the angular integration over $\phi_p$ and $\theta_p$ to get

\begin{multline}
\label{Disp_hom}
    D(k=0,\omega)= 1+\frac{16e_B^2}{\omega^2(2\pi)^2\hbar^3 \epsilon_0}\int dp \frac{p^2 \epsilon}{\hbar^2\omega^2-4\epsilon^2}
    \Big(1\\
    - \frac{p^2}{3\epsilon^2 } \bigg) 
    \bigg(
f-  \frac{\hbar^2\omega^2}{4\epsilon^2}
    \bigg) 
\end{multline}
Note here that the renormalization of the previous section has not been made in Eq. (\ref{Disp_hom}). This is on purpose, as repeating the renormalization procedure would obscure the computation of the imaginary contribution. Importantly, the pole contribution given rise to a wave damping term that is not subject to UV-divergences. Hence the imaginary part of the momentum integral can be computed independently of the renormalization. 

In the first term of the second parentheses we have dropped the subscript "e" on the background distribution $f$, as we only consider electrons (with immobile ions as a neutralizing background). The second term comes from the non-zero expectation value of the vacuum contribution. To handle the denominator in \cref{Disp_hom}, we separate the integral into the real principal value contribution and the imaginary pole contribution. The latter part will be evaluated at the resonant momenta $p=p_{res}$. Since $p_{res}$ never will approach infinity, we see that the issue of  renormalization will not affect this term. It is sufficient to note that the general procedure assures that it is the renormalized charge $e_r$ that must be used in the end. For the real part of the integral in \cref{Disp_hom}, the full momentum-space contributes, and hence the renormalization procedure of the previous section applies.

Focusing on the pole contribution,  it is convenient to factorize the denominator as follows
\begin{equation}
    D(k=0,\omega)= 1+\int dp \frac{G(p)}{(\hbar \omega-2\epsilon)(\hbar \omega+2\epsilon)}
\end{equation}
where $G(p)$ is the numerator of \cref{Disp_hom}. Without loss of generality, we may consider positive frequencies only, in which case the pole occurs at $\epsilon(p_{res})=\hbar \omega/2$. Changing integration variable from $p$ to $\epsilon$, evaluating the integral using the Landau contour, we deduce that the imaginary part of $D(k,\omega)$ is given by
\begin{multline}
{\rm Im}  (D)= \frac{G(\epsilon_{res})}{4 p_{res}} 
=-\frac{e_r^2 p_{res}}{6\pi \hbar^2 \omega  } \Big(1+\frac{2m^2}{\hbar^2\omega^2}\Big)\Big(f(p_{res})-1\Big) \label{Imhom}
\end{multline}


\begin{figure}
    \centering
    \includegraphics[width=\columnwidth]{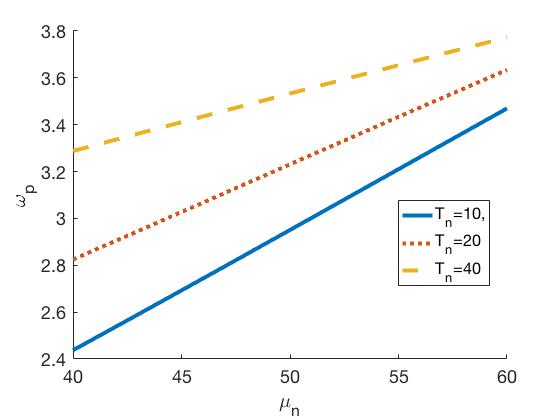}
    \caption{The plasma frequency is plotted as a function of the chemical potential $\mu_n$ for different temperatures. The solid line correspond to $T_n=10$, the dotted line $T_n=20$ and the dashed line $T_n=40$. }
    \label{Omega_p vs mu}
\end{figure}
\begin{figure}
    \centering
    \includegraphics[width=\columnwidth]{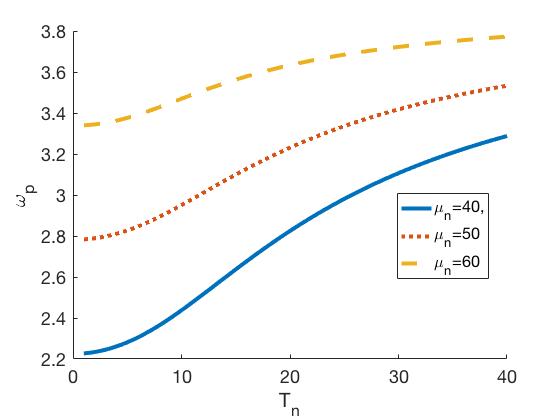}
    \caption{The plasma frequency is plotted as a function of the temperature $T_n$ for different chemical potentials. The solid line correspond to $\mu_n=40$, the dotted line $\mu_n=50$ and the dashed line $\mu_n=60$.}
    \label{Omega_p vs T}
\end{figure}
Until now, we have taken $f(p)$ as an isotropic but otherwise arbitrary background distribution function for the electrons. However, to be more specific, we assume the background to be in thermodynamic equilibrium, i.e. we pick a Fermi-Dirac distribution function 
\
\begin{equation}
\label{Fermi_Dirac}
    f(p)=\frac{1}{\exp{(\mu-\epsilon)}/k_BT}
\end{equation}
where $\mu$ is the chemical potential. 

As pointed out above, for wavelengths much longer than the Compton length, the relativistic Vlasov is a good approximation for the real frequency. Solving for the real frequency using the relativistic Vlasov equation in the homogeneous limit $k=0$, and then using Eq. (\ref{Imhom}), we can study the relative damping as a function of chemical potential and temperature. As a prerequisite, in Fig 2, the real part of the frequency $\omega_r=\omega_p$ is shown as a function of the chemical potential for different temperatures. Since the plasma density increases with chemical potential, naturally $\omega_r$ is a strictly increasing function of $\mu$. Similarly, in Fig 3, for three different chemical potentials, $\omega_r$ is an increasing function of $T$. We note that since a pair-creation resonance is assumed to exist, we must have a sufficiently high plasma frequency to start with, i.e. there are restrictions on the choice of $\mu$ and $T$. For the regimes plotted in Figs. 2 and 3, the pair-creation condition is fulfilled, however.  

\begin{figure}
    \centering
    \includegraphics[width=\columnwidth]{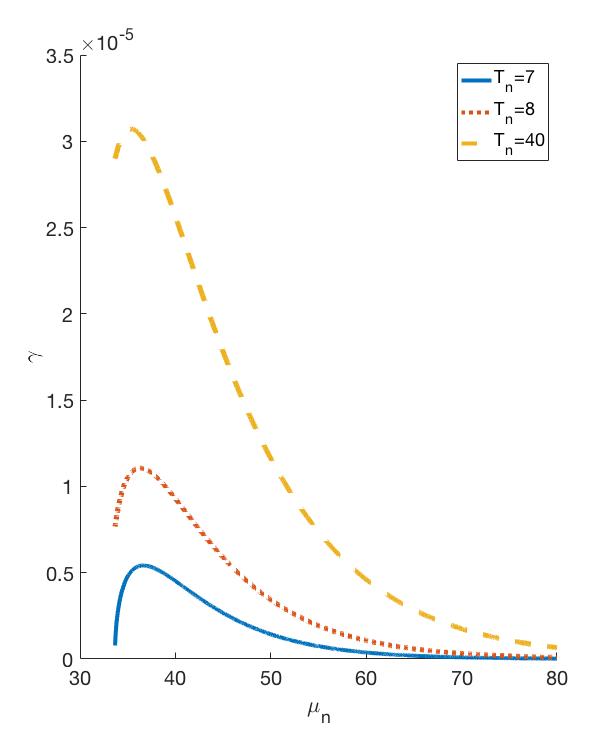}
\caption{The damping $\gamma$ as a function of the normalized chemical potential $\mu_n$ for different temperatures $T_n$. The solid curve is for $T_n=7$, the dotted line for $T_n=8$ and the dashed line for $T_n=10$.}
    \label{Imeps vs mu}
\end{figure}

Next we move on to the damping rate. In \cref{Imeps vs mu}, the relative damping  $\gamma=\omega_i/\omega_r$ is plotted as a function of the normalized chemical potential $\mu_n=\mu/m$ for different normalized temperatures $T_n=T/m$. As can be noted, the damping has a peak-value around $\mu_n \sim 40$, where the exact peak is slightly different for different temperatures. For larger chemical potentials (higher densities of the background plasma) the relative damping is suppressed.
The reasons for this suppression is as follows: For very large $\mu$, the resonant momentum $p_{res}$ is located at the low-energy end of the distribution function, where the vacuum and particle distribution tend to cancel approximately, and hence we get a weaker damping. Similarly, one can notice in \cref{Imeps vs mu} that higher temperatures gives a stronger damping. The reason is that for $T\ll \mu $, the contributions from the plasma and vacuum to the integrand tend to cancel close to the resonant momentum. This can be verified by looking at the  second parentheses in \cref{Disp_hom}. If $T\ll \mu$ we have $f \approx 1\approx \hbar \omega/2\epsilon$. For higher temperatures, we have less degenerate states and the background function $f$ is not close to a step-function of energy. Thus the approximate cancellation between the vacuum and the plasma contribution close to the resonance is broken.

\begin{figure}
    \centering
    \includegraphics[width=\columnwidth]{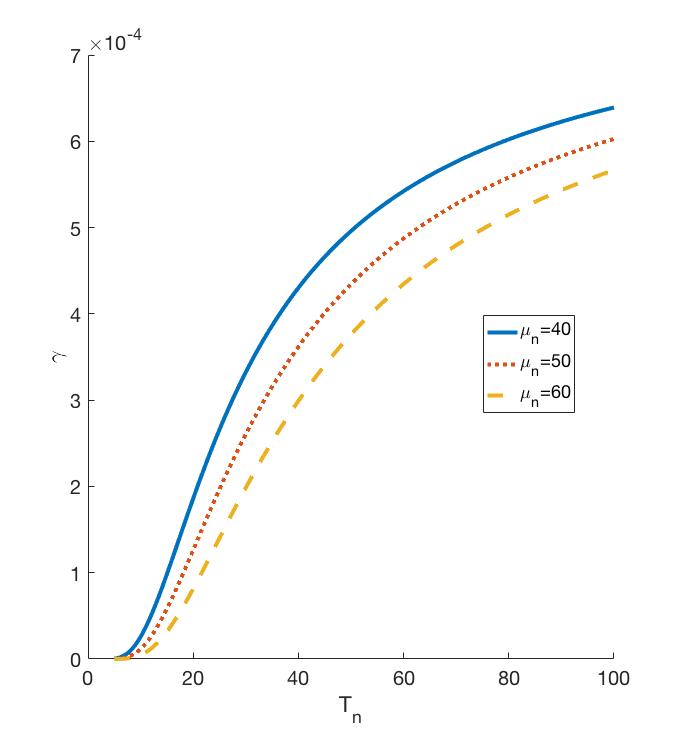}
\caption{The damping $\gamma$ as a function of the normalized temperature $T_n$ for different chemical potentials $\mu_n$. The solid line correspond to $\mu_n=40$, the dotted line $\mu_n=50$ and the dashed line $\mu_n=60$.}
    \label{Imeps vs T}
\end{figure}
 The temperature-dependence of the damping is seen more clearly in \cref{Imeps vs T}. The damping is much increased for higher temperatures, and this is most pronounced for $T<\mu$ since the produced pairs have modest gamma-factors. For $T\geq \mu $ the produced pairs are created with larger gamma-factors and hence the pair production rate and the corresponding wave-damping is diminished.

\subsection{General k-dependence}
In this subsection we will study the damping as described by \cref{Dispersion relation} for a general $k$, except that we assume $k$ to be well below the Compton wavenumber.   Thus the real part of the dispersion relation can still be obtained using the relativistic Vlasov eqution. However, the imaginary part becomes more complex to evaluate when a non-zero $k$ is introduced. Specifically, the resonant momentum $p_{res}$ is modified to 
\begin{equation}
    p_{res}= \frac{1}{2}\sqrt{\frac{ \hbar^2 (\omega^2-k^2) -4m^2}{1-\frac{k^2}{\omega^2}\cos^2{\theta_p} }} \label{presk}
\end{equation}
where $p_z=p cos\theta_p $.
A finite wave-vector $k$ will make the requirement for a non-zero damping more strict, i.e. we need a higher plasma density in order for \cref{presk} to be fulfilled for a real value of $p_{res}$. Physically, in order to produce a pair from one plasmon, we have to fulfill the conservation of energy and momentum. Thus when the wave quanta carries momentum $\hbar k$ in addition to the  energy $\hbar \omega$, electron-positron pairs with minimum energy (zero momentum) cannot be created, as the created pairs must absorb a finite momentum. As a result, a higher plasma density is required for pair-creation damping to be possible.

\begin{figure}
    \centering
    \includegraphics[width=\columnwidth]{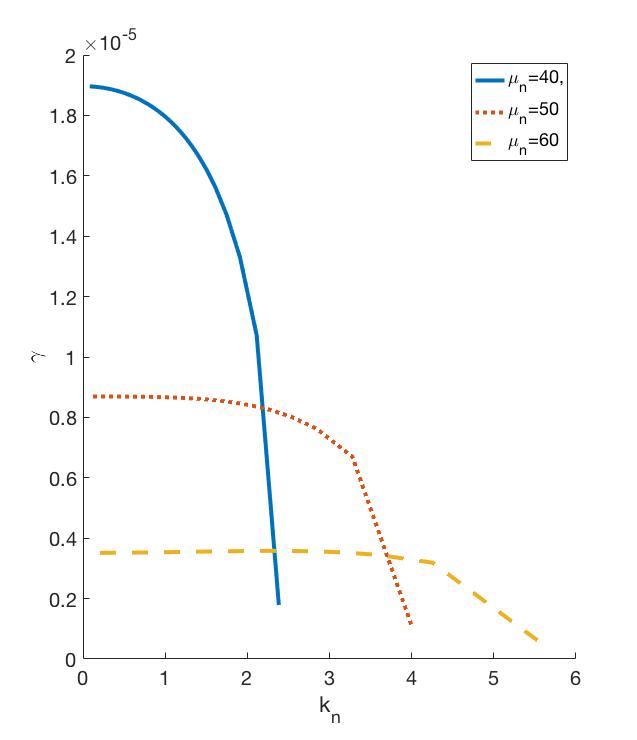}
\caption{The damping $\gamma$ as a function of the normalized wave-vector $k_n$ for different chemical potentials $\mu_n$ with $T_n=10$. The solid line correspond to $\mu_n=40$, the dotted line $\mu_n=50$ and the dashed line $\mu_n=60$.}
    \label{Imeps vs k mu}
\end{figure}

\begin{figure}
    \centering
    \includegraphics[width=\columnwidth]{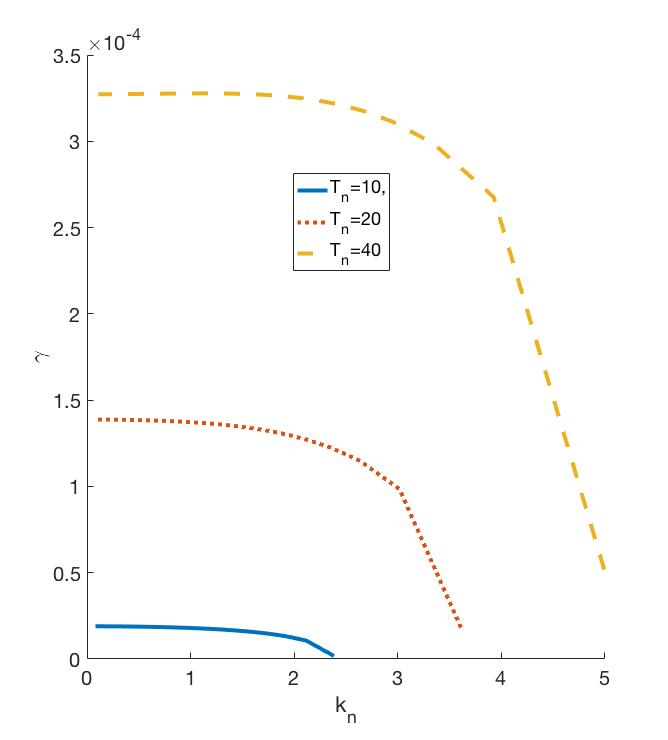}
\caption{The damping $\gamma$ as a function of the normalized wave-vector $k_n$ for different temperatures $T_n$ with $\mu_n=40$. The solid line correspond to $T_n=10$, the dotted line $T_n=20$ and the dashed line for $T_n=40$}
    \label{Imeps vs k T}
\end{figure}

Next, we solve for the relative damping $\gamma $ numerically applying the same procedure as in the proceeding section, using the relativistic Vlasov equation for the real frequency, still taking the background distribution to be given by \cref{Fermi_Dirac}. 
The relative damping $\gamma $ is plotted as a function of the normalized wave-vector $k_n=\frac{\hbar k}{m}$ for different chemical potentials $\mu_n$ having the temperature $T_n=10$ in \cref{Imeps vs k mu}. As one would expect, the damping is suppressed for larger $k_n$ since the pair-creation condition becomes increasingly difficult to fulfill. To fulfill the resonant condition \cref{presk}, harder photons needs to be absorbed in order to create pairs with larger gamma-factors. However, there is a limit for how large $k_n$ can be without the momentum resonance disappearing. We denote this value with $k_{max}$. This value can be obtained from \cref{presk}
\begin{equation}
    k_{max}= \sqrt{\omega^2-4\omega_{c}^2}. \label{kmax}
\end{equation}
Whether or not a real value of $k_{max}$ fulfilling \cref{kmax} is possible, depends on the plasma frequency (whether is is high enough), which in turn depends on the temperature and the density of the plasma, as computed in \cref{Omega_p vs mu,Omega_p vs T}. This can be confirmed by studying the dependence of the damping rate in \cref{Imeps vs k mu,Imeps vs k T} where larger $\mu_n$ and $T_n$ gives a higher value of $k_{max}$.  In \cref{Imeps vs k mu} we note that for $T=10$, a smaller chemical potential give larger damping for $k=0$, since a more degenerate system tends to suppress the damping. However, this behavior is turned around for larger $k$, since a high plasma frequency is needed for damping to be possible, as reflected by the condition (\ref{kmax}). The combined dependence on temperature and on wavenumber is more straightforward, as shown in \cref{Imeps vs k T}, as the damping generally is increasing with $T$ and decreasing with $k$.

\subsection{Comparison with numerical result}
Next, in order to confirm the analytical calculations of the preceding sections, we want to solve \cref{PDE_System} numerically. To simplify the numerical calculation of \cref{PDE_System}, we study the homogeneous limit, i.e. we drop the $z$-dependence in the system (this corresponds to the $k \rightarrow 0$-limit in the linear case). In the homogeneous limit, the charge density $v_0$ remains zero and in terms of the $\chi$-variables $\chi_4 =0$. In order to
 simplify the numerical calculations, 
  we define new variables $\Tilde{\chi}_i(\bp,t)$ as the deviation from the vacuum state, i.e. we let
 
 \begin{equation}
     \Tilde{\chi}_i(\bp,t)=\chi_i(\bp,t)- \chi_{i {\rm vac}}(\bp).
    \end{equation}
    where $\chi_{ivac}(\bp)$ is given by the vacuum-part of \cref{Initial_values}. 
    Note that $\Tilde{\chi}_{3}={\chi}_{3}$. 
 Secondly, we switch to canonical momentum
 \begin{equation}
     q=p_z+eA
 \end{equation}
 and use the Gauge where the scalar potential is zero such that $E=-\partial A/\partial t$. With canonical momentum $q$ as one of the independent variables, the operator $D_t$ simplifies to $D_t=\partial/\partial t$. Our final step is to switch to dimensionless variables.  The normalized variables are given by, $t_n= (m/\hbar) t$, $q_n=q/m$, $p_{n\perp}=p_{\perp}/m$, $E_n=E/E_{cr}$, $A_n=eA/m$, where $E_{cr}=m^2/e\hbar$ is the critical electric field.. We note that the DHW-functions are already normalized. With the above steps performed, the equations to be solved numerically read: 
\begin{align}
\label{PDE_System4}
    \frac{\partial \Tilde{ \chi}_1}{\partial t} (q,p_{\bot},t)&= 2\epsilon_{\bot} \Tilde{\chi}_3 + 2E\frac{\epsilon_{\bot}^2}{\epsilon^3} \notag\\
      \frac{\partial  \Tilde{\chi}_2}{\partial t} (q,p_{\bot},t )&= -2(q-A)\Tilde{\chi}_3-2(q-A)E\frac{\epsilon_{\bot}}{\epsilon^3}\notag\\
      \frac{\partial \Tilde{\chi}_3}{\partial t} (q,p_{\bot},t)  &=-2\epsilon_{\bot} \Tilde{\chi}_1 +2(q-A)\Tilde{\chi}_2
      \end{align}
      
      with Ampere´s law
      \begin{equation}
\label{Ampers_law2}
\frac{\partial E}{\partial t}=- \eta \int \chi_1 d^3p
\end{equation}

where the dimensionless factor is $\eta=\alpha/(2\pi^2)\approx0.3697\times10^{-3}$. For notational convenience, we have dropped the subscript $n$. 
We solve this system for $E=0.01 E_{cr}$, which is sufficiently low to suppress nonlinear effects. For the initial values of the $ \Tilde{\chi}_{i}$-variables, we use the plasma part of the background given in \cref{Initial_values} since $\Tilde{\chi}_i$ is defined as the deviation from the vacuum-state. The background distribution function of the plasma is defined in \cref{Fermi_Dirac}. Finally, the numerical system is solved using a leapfrog method. As expected, the electric field evolves in accordance with linear theory, i.e. we have harmonic oscillations, superimposed with a small exponential damping. The real part of the frequency is well approximated by the relativistic Vlasov equation, in accordance with the analytical theory.  With the basic features confirmed, we want to compare the dependence of the damping rate on the background parameters in the analytical and numerical calculations. For this purpose, we fit the numerical profile to the theoretical function \begin{equation}
    E(t)=E_0e^{-i\omega t}=E_0e^{-i\omega_r }e^{\omega_it}
\end{equation}
where we let $\omega=\omega_r+i\omega_i$. The damping-term can be rewritten as
\begin{equation}
    e^{\omega_it}=e^{\omega_i 2\pi N/\omega_r}=e^{2\pi N \gamma}
\end{equation}
where $N$ is the number of plasma periods that we follow the evolution numerically. Computing the numerical drop in amplitude after N plasma periods, the numerical expression for the relative damping $\gamma$ is 
\begin{equation}
    \gamma =\frac{1}{2\pi N} \ln{ \bigg( \frac{E(t=NT)}{E(t=0)}\bigg) }
\end{equation}

\begin{figure}
    \centering
\includegraphics[width=\columnwidth]{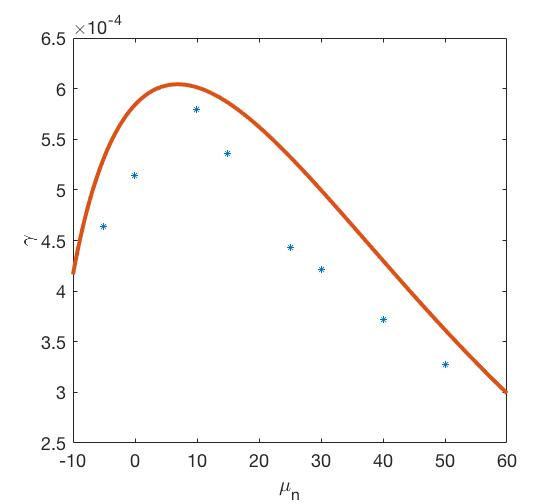}
    \caption{The relative damping $\gamma$ is plotted as a function of the chemical potential $\mu_n$ for $T_n=40$. Here the solid line represents the analytical result while the dots are from the numerical calculation.}
    \label{gamma_vs_mun}
\end{figure}

The above expression for the relative damping $\gamma$ is plotted as a function of the chemical potential $\mu_n$ in \cref{gamma_vs_mun} together with the corresponding values from the analytical solution. It is clear that the numerical points follows the analytical curve and hence there is a quantitative agreement, although the precision is far from perfect. However, with the numerical calculation made on a laptop, the computation is quite time-consuming and must be made with a fairly modest resolution.  Decreasing the step length (time step, as well as perpendicular and parallel momentum steps), the quantitative agreement is improved. However, even the present agreement is sufficient to effectively confirm the analytical findings.

\begin{figure}
    \centering
\includegraphics[width=\columnwidth]{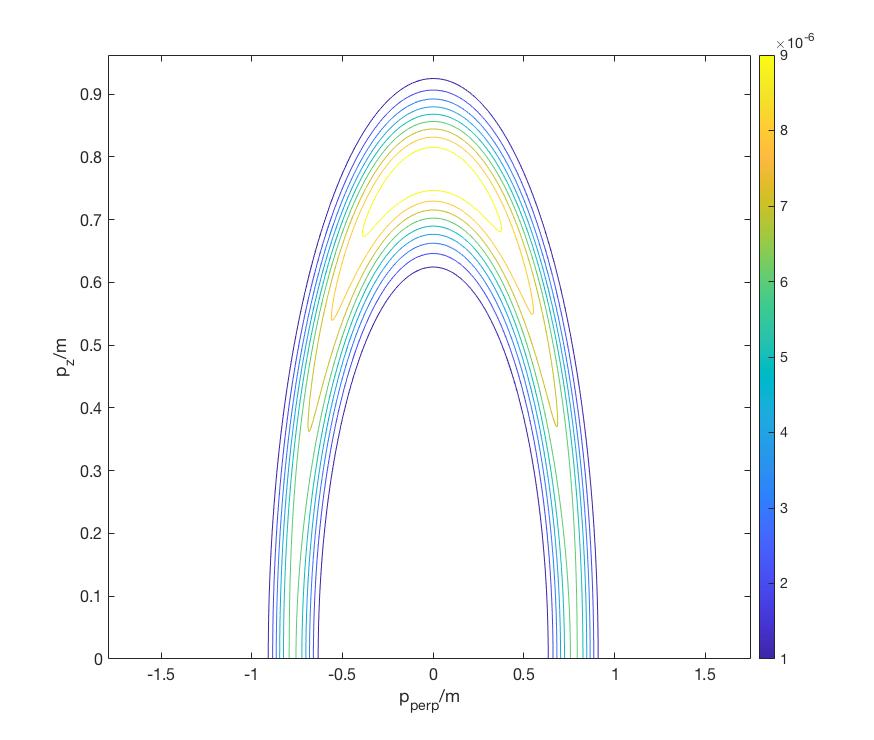}
    \caption{A contour plot of the produced pairs in the momentum-space.}
    \label{Contour_plot}
\end{figure}

Before concluding this section, let us study the loss mechanism in some detail. Generally for linear pair creation, the wave energy decreases and the particle energy increases. We can use this simple fact to demonstrate where the electron-positron pairs are located in momentum-space. Based on \cref{Conservation_Energy} and \cref{Particle_energy}. one can interpret $n(\bp,t)$ as the particle density in momentum space, see \cite{PRE21} for more details. 
The electron-positron density in the momentum-space can be expressed as
\begin{equation}
    n_{pair}(\bp)=n(\bp,t)-n(\bp,t=0)
\end{equation}
where we subtracted the initial particle density of the plasma $n(\bp,t=0)$. Using the same initial data as before, a contour plot of $n_{pair}(\bp)$ can be found in \cref{Contour_plot}. Here the contour plot is made at a time $t$ that corresponds to several wave periods, such that distribution of the pairs in momentum-space is evident.

The main feature of importance in \cref{Contour_plot} is the demonstration that \cref{Resonant_mom} is fulfilled, as the contour curves of particle production are centered around the resonant momentum. The extra information provided by \cref{Contour_plot}, that goes beyond the analytical calculation made in section IV, is the width of the momentum resonance.     
\section{Summary and Conclusion}
In the present paper, we have used the DHW-formalism to study linearized Lamgmuir waves in very dense plasmas. The vacuum contribution to the current and mass density leads to ultra-violet divergences, that must be handled using a renormalization. After removing the divergent terms, a contribution from the vacuum polarization still remains. While this contribution is negligible for most purposes, it is nevertheless interesting to confirm that the contribution found here agrees with an expression derived using independent methods \cite{Mameyev-1981}. 

For a relativistic Fermi velocity and wavelengths much longer than the Compton length, the quantum effect have a relatively small effect on the real part of the wave-frequency. That is, after picking a suitable background distribution (possibly degenerate) the classical (but relativistic) Vlasov equation gives an accurate approximation for $\omega_r$. However, pair-creation damping is a main new feature of the DHW-formalism, for densities high enough to fulfill the pair-creation condition. The magnitude of this process depends on temperature, density, and wave-number. The scaling with these parameters have been analyzed, and physical interpretations of the results have been given. 

In this paper, we have only studied damping due to pair-creation. In principle, however,just as for regular Landau damping, the integrand can change sign at the resonance region. This will happen for an electron-positron plasma of sufficient density, in which case we can have wave growth due to collective annihilation, which can occur at a much faster rate than ordinary two-particle (collisional) annihilation. In practice, however, this would require a very dense electron-positron plasma, which is at least partially degenerate. As far as we know, no such plasmas exist in the universe, and hence we have not analyzed this case, although it is a theoretical possibility.     

While our results have been limited to electrostatic wave modes in unmagnetized plasmas, pair-creation damping should be possible for any plasma wave modes fulfilling the pair creation condition, which read $\hbar^2(\omega^2-k^2)>4m^2c^4$ in general. Unless the plasma frequency or the electron cyclotron frequency is very high, this condition will typically not be fulfilled, however. Generally very high density plasmas is required for pair-creation damping, e.g. like in white dwarf stars or neutron stars, and in the early universe. Moreover, it can be noted that in weakly nonlinear theories, processes involving multiple wave-quantas are possible \cite{Brodin-2017}, in which case the pair-creation condition may be relaxed somewhat. In particular, by analogy with the results from Ref. \cite{Brodin-2017}, we expect the pair-creation condition for {\it three plasmon processes} to read $9\hbar^2(\omega^2-k^2)>4m^2c^4$, i.e. the condition for the minimum plasma frequency is relaxed by a factor of three. Studies of such processes, however, is beyond the scope of the present paper. 
\appendix
\section{Validity of the DHW-equations}
\label{Validity of DHW}
The derivation of the DHW-equations makes use of the Hartree (mean-field) approximation , where electromagnetic field is treated as a non-quantized field, and thus all quantum fluctuations of the field is dropped. When discussing the applicability of the mean field approximation, Ref. \cite{Birula} (where the system was first derived) has argued that it likely accurate for the case of strong electromagnetic  fields, in case the temporal variations are not too rapid. As far as we know, a more precise assessment of the applicability has not been made. While the conditions of "strong field" and slow temporal variations makes sense if one is primarily concerned with low density physics (i.e. problems like particle creation from vacuum due to the Schwinger mechanism), it is not necessarily fitting for the case of a high density plasma. 

To illustrate the role played by the plasma, let us define the parameter $N=nL^3$, where $n$ is the electron number density, $L^3$ is the characteristic volume occupied by a field quanta, and $N$ is the characteristic number of electrons simultaneously interacting with a field quanta. When the electromagnetic field is dominated by a single wavelength $\lambda$, it is safe to deduce
\begin{equation}
N=nL^3>n\lambda^3=\frac{\omega_p^2}{k^2c^2}\,4\pi^2 \frac{\lambda}{r_c}.
\end{equation}
Here we have introduced the square of the plasma frequency $\omega_p^2=ne^2/\epsilon m$, and the classical electron radius, $r_c=(1/4\pi\epsilon_0)(e^2/mc^2)\approx2\times10^{-15}{\rm m}$. Limiting ourselves to unmagnetized plasmas, and cases where the plasma is dense enough to significantly alter the vacuum dispersion relation, we will have $(\omega_p^2/k^2c^2)\gtrapprox 1$, independently whether we consider electromagnetic or electrostatic wave modes. For such cases, even for very short wavelengths (of the order of the Compton wave length), we will have
\begin{equation}
    N\gtrapprox3\times 10^4\gg1,
\end{equation}
such that any wave quanta will simultaneously interact with a very large number of electrons. This strongly suggests that the matter response to the fields will be a collective one, such that the mean field approximation will be appropriate. Although this arguments extends the applicability of the DHW-formalism beyond the Compton frequency, note that it only applies in case the plasma density is high enough to make the plasma frequency equally high.

\end{document}